\documentclass[11pt,twocolumn,a4paper]{article}
\usepackage[utf8]{inputenc}
\usepackage[T1]{fontenc}
\usepackage{amsmath}
\usepackage{amsfonts}
\usepackage{amssymb}
\usepackage{fullpage}
\usepackage{bbold}
\usepackage{xcolor}
\usepackage{color}
\usepackage{xspace}
\usepackage{lscape}
\usepackage{tabularx}
\usepackage{url}
\usepackage{natbib}
\usepackage[font={small}]{caption}
\usepackage[affil-it]{authblk}
\usepackage{mathtools}
\usepackage{float}
\usepackage{hyperref}
\usepackage{longtable}
\usepackage{float}
\usepackage[hmargin={0.7in,0.7in},vmargin={0.8in,0.8in}]{geometry}
\usepackage[space]{grffile}
\usepackage{caption}
\usepackage{subcaption}
\usepackage{graphicx}
\usepackage{epstopdf} 
\usepackage{algpseudocode}
\usepackage{booktabs}
\usepackage{footmisc}
\newcommand{\possessivecite}[1]{\citeauthor{#1}'s \citeyearpar{#1}}
\usepackage{abstract}

\title{Long-run dynamics of the U.S. patent classification system
\thanks{This paper reuses material from an unpublished chapter of the first author's PhD thesis at UNU-MERIT, Maastricht University. This work was supported by the National Research Foundation of Korea (NRF) funded by the Korean Government [Grants No. NRF-2017R1A2B3006930 (D.K.)], and the European Commission project FP7-ICT-2013-611272 (GROWTHCOM) (F.L.). We also acknowledge support from Partners for a New Economy, the London Institute for Mathematical Sciences, the Institute for New Economic Thinking at the Oxford Martin School, and the Oxford Martin School Programme on Technological and Economic Change. We have benefited from excellent comments from two anonymous referees and many colleagues, including Jeff Alstott, Yuki Asano, Mariano Beguerisse D\'{i}az, J. Doyne Farmer, Marco Pangallo, Emanuele Pugliese, Giorgio Triulzi and Vilhelm Verendel. We are also grateful to Diana Greenwald and the USPTO for helping us locating data sources. All remaining errors are ours. Contacts: francois.lafond@inet.ox.ac.uk, daniel.youngho.kim@gmail.com}
}

\author[1,2,3]{Fran\c{c}ois Lafond}
\author[4]{Daniel Kim}

\affil[1]{Institute for New Economic Thinking at the Oxford Martin School, University of Oxford}
\affil[2]{Smith School for Enterprise and the Environment, University of Oxford}
\affil[3]{Oxford Martin School Programme on Technological and Economic Change, University of Oxford}
\affil[4]{Natural Science Research Institute, Korea Advanced Institute of Science and Technology%, Daejeon 34141, Republic of Korea
}

\date{\today}

\begin{document}

\twocolumn[
 \begin{@twocolumnfalse}
    \maketitle
\begin{abstract}

Almost by definition, radical innovations create a need to revise existing classification systems. In this paper, we argue that classification system changes and patent reclassification are common and reveal interesting information about technological evolution. To support our argument, we present three sets of findings regarding classification volatility in the U.S. patent classification system. First, we study the evolution of the number of distinct classes. Reconstructed time series based on the current classification scheme are very different from historical data. This suggests that using the current classification to analyze the past produces a distorted view of the evolution of the system. Second, we study the relative sizes of classes. The size distribution is exponential so classes are of quite different sizes, but the largest classes are not necessarily the oldest. To explain this pattern with a simple stochastic growth model, we introduce the assumption that classes have a regular chance to be split. Third, we study reclassification. The share of patents that are in a different class now than they were at birth can be quite high. Reclassification mostly occurs across classes belonging to the same 1-digit NBER category, but not always. We also document that reclassified patents tend to be more cited than non-reclassified ones, even after controlling for grant year and class of origin.

Keywords: patents, classification, reclassification.

JEL codes: O30, O39.

\end{abstract}
\vspace{1cm}

  \end{@twocolumnfalse}
]
\saythanks

\section{Introduction}

The U.S. patent system contains around 10 million patents classified in about 500 main classes. However, some classes are much larger than others, some classes are much older than others, and more importantly none of these classes can be thought of as a once-and-for-all well defined entity. Due to its important legal role, the U.S. Patent and Trademark Office (USPTO) has constantly devoted resources to improve the classification of inventions, so that the classification system has greatly evolved over time, reflecting contemporaneous technological evolution. Classifications evolve because new classes are created but also because existing classes are abolished, merged and split. In fact, all current classes in 2015 have been established in the U.S. Patent Classification System (USPCS) after 1899, even though the first patent was granted in 1790 and the first classification system was created in 1829-1830. To give just another example, out of all patents granted in 1976, 40\% are in a different main class now than they were in 1976.

To maintain the best possible level of searchability, the USPTO reclassifies patents so that at every single moment in time the patents are classified according to a coherent, up-to-date taxonomy. The downside of this is that the current classification is not meant to reflect the historical description of technological evolution as it unfolded. In other words, while the classification system provides a consistent classification of all the patents, this consistency is not time invariant. Observers at different points in time have a different idea of what is a consistent classification of the past, even when classifying the same set of past patents.
In this paper, we focus on the historical evolution of the U.S. patent classification. We present three sets of findings.

First we study the evolution of the number of distinct classes, contrasting current and historical classification systems. Recent studies \citep{strumsky2012using,strumsky2015identifying,youn2015invention} have shown that it is possible to reconstruct the long-run evolution of the number of subclasses using the current classification system. This allowed them to obtain interesting results on the types of recombinations and on the relative rates of introduction of new subclasses and new combinations. An alternative way to count the number of distinct categories is to go back to the archives and check how many classes did actually exist at different points in the past. We found important differences between the historical and reconstructed evolution of the classification system. In particular, we find that historically the growth of the number of distinct classes has been more or less linear, with about two and a half classes added per year. By contrast, the reconstructed evolution \--- which considers how many current classes are needed to classify all patents granted before a given date \--- suggests a different pattern with most classes created in the 19$^{th}$ century and a slowdown in the rate of introduction of novel classes afterwards. Similarly, using the historical classes we find that the relationship between the number of classes and the number of patents is compatible with Heaps' law, a power law scaling of the number of categories with the number of items, originally observed between the number of different words and the total number of words in a text \citep{heaps1978information}. Using the reconstructed evolution Heaps' law does not hold over the long run.

Knowing the number of distinct classes, the next question is about their growth and relative size (in terms of the number of patents). Thus our second set of findings concerns the size distribution of classes. We find that it is exponential, confirming a result of \citet{carnabuci2013distribution} on a much more restricted sub-sample. We also find that there is no clear relationship between the size and the age of classes, which rules out an explanation of the exponential distribution in terms of simple stochastic growth models in which classes are created once and for all.

Third, we hypothesize that new technology fields and radical innovations tend to be associated with a higher reclassification activity. This suggests that the history of reclassification contains interesting information on the most transformative innovations. Our work here is related to \citet{wang2016technological} who study how a range of metrics (claims, references, extensions, etc.) correlate with reclassification for 3 million utility patents since 1994. We used the data since 1976, for which we observe the class of origin and the citations statistics. It appears that reclassified patents are more cited than non-reclassified patents. We also construct a reclassification flow diagram, with aggregation at the level of NBER patent categories \citep{hall2001nber}. This reveals that a non-negligible share of patents are reclassified across NBER categories. We find that patents in ``Computers'' and in ``Electronics'' are often reclassified in other NBER categories, which is not the case of other categories such as ``Drugs''. We then discuss three examples of new classes (Fabric, Combinatorial Chemistry and Artificial Intelligence).

Finally, we argue that it is not possible to explain the observed patterns without accounting for reclassification. We develop a simple model in which classes grow according to preferential attachment but have a probability of being split. The model's only inputs are the number of patents and classes in 2015 and the Heaps' law exponent. Despite this extreme parsimony, the model is able to reproduce i) the historical and reconstructed patterns of growth of the number of classes, ii) the size distribution and (partially) the lack of age-size relationship, and iii) the time evolution of the reclassification rates.

The empirical evidence that we present and the assumptions we need to make for the model make it clear that the USPCS has evolved considerably and it is hardly possible to think of patent classes as technological domains with a stable definition. The classification system cannot be well understood as a system in which categories are created once-and-for-all and accumulate patents over time. Instead, it is better understood as a system that is constantly re-organized. Because of this, using the current classification system to study a set of older patents is akin to looking at the past with today's glasses. In this paper, we not only show the differences between the historical and reconstructed reality, but we also explain how these differences emerged.

The paper is organized as follows. Section \ref{section:motiv} details our motivation, gives some background on categorization and reviews the literature on technological categories. Section \ref{section:data} describes the USPCS and our data sources. Section \ref{section:Nclass} presents our results on the evolution of the number of classes. Section \ref{section:sizedist} discusses the size distribution of classes. Section \ref{section:reclass} presents our results on reclassification since 1976. Section \ref{section:model} presents a model that reproduces the main empirical patterns discovered in the previous sections. The last section discusses the results, motivates further research and concludes.

\section{Why is studying classification systems important?}
\label{section:motiv}

Classification systems are pervasive because they are extremely useful. At a fundamental level, categorization is at the basis of pattern recognition, learning, and sense-making. Producing a discourse regarding technologies and their evolution is no exception. As a matter of fact, theoretical and \emph{a fortiori} empirical studies almost always rely on some sort of grouping \--- or aim at defining one. 

Historically, the interest in technology classifications has been mostly driven by the need to match technological and industrial activities \citep{schmookler1966invention,scherer1984using,verspagen1997measuring}. Since patented technologies are classified according to their function, not their industry of use or origin, this problem is particularly difficult. Clearly, a good understanding of both industry and patent classification systems is crucial to build a good crosswalk. Here we highlight the need to acknowledge that both classification systems \emph{change}. For this reason our results give a strong justification for automated, probabilistic, data-driven approaches to the construction of concordance tables such as the recent proposal by \citet{lybbert2014getting} which essentially works by looking for keywords of industry definitions in patents to construct technology-industry tables.

With the rise of interest in innovation itself many studies have used existing patent classifications to study spillovers across technology domains, generally considering classification as static. For instance 
\citet{kutz2004examining} studied the growth and distribution of patent classes since 1976;  \citet{leydesdorff2008patent}, \citet{antonelli2010recombinant}, \citet{strumsky2012using} and \citet{youn2015invention} studied co-classification patterns; and \citet{caminati2010pattern} and \citet{acemoglu2016innovation} studied the patterns of citations across USPCS or NBER technology classes. Similarly, technological classification systems are used to estimate technological distance, typically between firms or inventors in the ``technology space'' based on the classification of their patent portfolio \citep{breschi2003knowledge,nooteboom2007optimal,aharonson2016mapping,alstott2016mapping}. Additional methodological contributions include \citet{benner2008close}, who have pointed out that using all the codes listed on patents increases the sample size and thus reduces bias in measuring proximity, and \citet{mcnamee2013can} who argues for using the hierarchical structure of the classification system\footnote{In a related context (how professional diversity scales with city size), \citet{bettencourt2014professional} and \citet{youn2016scaling} exploited the different layers of industry and occupation classifications systems to identify resolution-independent quantities. Measuring diversity depends on which layer of the classification system one uses, but in such a way that the infinite resolution limit (deepest classification layer) exists and can be used to characterise universal quantities.}.

In spite of this wide use of the current patent classification system, there have been no quantitative studies of the historical evolution of the system apart from the counts of the number of distinct classes by \citet{bailey1946history} and \citet{stafford1952rate}, which we update here. Recently though, \citet{strumsky2012using} originated a renewed interest in patent classification by arguing that the classification of patents in multiple fields is indicative of knowledge recombination. Using the complete record of US patents classified according to the current classification system, \citet{youn2015invention} studied the subclasses (``technology codes''). They found that the number of subclasses used up to a given year is proportional to the cumulative number of patents until about 1870, but grew less and less fast afterwards. Remarkably, however, this slowdown in the ``introduction'' of new subclasses does not apply to new \emph{combinations} of subclasses. \citet{youn2015invention} found that the number of combinations has been consistently equal to 60\% of the number of patents. This finding confirms \possessivecite{strumsky2012using} argument that patent classifications contain useful information to understand technological change over the long-run. Furthermore, the detailed study of combinations can reveal the degree of novelty of specific patents \citep{strumsky2015identifying,kim2016technological}. 

Besides their use for simplifying the analysis and creating crosswalks, technology taxonomies are also interesting \emph{per se}. A particularly interesting endeavour would be to construct systematic technology phylogenies showing how a technology descends from others \citep{basalla1988evolution,sole2013evolutionary} (for specific examples, see \citet{temkin2007phylogenetics} for cornets and  \citet{valverde2015punctuated} for programming languages). 

But categories are not simply useful to describe reality, they are often used to \emph{construct} it \citep{foucault1966mots}. When categories are created as nouns, they can have a predicate and become a subject. As a result, classification systems are institutions that allow agents to coordinate and agree on how things should be called and on where boundaries should be drawn. Furthermore, classification systems may create a feedback on the system it describes, for instance by legitimizing the items that it classifies or more simply by biasing which items are found through search and reused in recombination to create other items. Categorization thus affects the future evolution of the items and their relation (boundaries) with other items. Along this line of argument, the process of categorization is performative. In summary, data on the evolution of technological classification systems provides a window on how society understands its technological artefacts and legitimizes them through the process of categorization. According to \citet{latour2005reassembling}, social scientists should not over impose their own categories over the actors that they analyze. Instead a researcher should follow the actors and see how they create categories themselves.

\citet{nelson2006perspectives} described technological evolution as the co-evolution of a body of practice and a body of understanding. The role of the body of understanding is to ``rationalize'' the practice. According to him this distinction has important implications for understanding evolutionary dynamics, since each body has its own selection criteria. Our argument here is that the evolution of the USPCS reflects how the beliefs of the community of technologists about the mesoscale structure of technological systems coevolves with technological advancements. We consider patent categorization as a process of codification of an understanding concerning the technological system. To see why studying patent categories goes beyond studying patents, it is useful to remember that examiners and applicants do not need to prove that a technology improves our \emph{understanding} of a natural phenomenon; they simply need to show that a device or process is novel and effective at solving a problem. However, to establish a new class, it is necessary to agree that bringing together inventions under this new header actually improves understanding, and thus searchability of the patent system. In that sense we believe that the dynamics of patent classes constitute a window on the ``community of technologists''.\footnote{
Patent officers are generally highly skilled workers. Besides anecdotal evidence on particularly smart patent examiners (Albert Einstein), patent officers are generally highly qualified (often PhDs). That said, \citet{rotkin1999history} mention that classification work was not particularly attractive and that the Classification division had difficulties attracting volunteers. More recently \citet{paradise2012claiming} eludes to ``high turnover, less than ideal wages and heavy workloads''. There is an emerging literature on patent officers' biases and incentives \citep{cockburn2003all,schuett2013patent} but it is focused on the decision to grant the patent. Little is known about biases in classification.
}
Since classification systems are designed to optimize search, they reflect how search takes place which in turn is indicative of what thought processes are in place. These routines are an integral part of normal problem-solving within a paradigm. As a result, classification systems must be affected by paradigm-switching radical innovations. As noted by e.g. \citet{pavitt1985patent} and \citet{hicks2011structural}, a new technology which fits perfectly in the existing classification scheme may be considered an incremental innovation, as opposed to a radical innovation which challenges existing schemes. A direct consequence is that the historical evolution of the classification system contains a great deal of information on technological change beyond the information contained in the patents\footnote{In labor economics, some studies have exploited classification system changes. \citet{xiang2005new} finds that new goods, as measured by changes to the SIC system, have a higher skill intensity than existing goods. \citet{lin2011technological} and \citet{berger2015industrial} used changes in the index of industries and the dictionary of occupational titles to evaluate new work at the city level.}. We now describe our attempt at reconstructing the dynamics of the U.S. patent classification system.

\section{The data: the USPCS}
\label{section:data}

We chose the USPCS for several reasons. First of all, we chose a patent system, because of our interest in technological evolution but also because due to their important legal role patent systems benefit from resources necessary to be maintained up to date. Among the patent classification systems, the USPCS is the oldest still in use (as of couple of years ago) \citep{wolter2012takes}. It is also fairly well documented, and in english. Moreover, additional files are available: citation files, digitized text files of the original patents from which to get the classification at birth, files on current classification, etc. Finally, it is one of the most if not the most used patent classification system in studies of innovation and technological change.
The major drawback of this choice is that the USPCS is now discontinued. This means that the latest years may include a classificatory dynamics that anticipate the transition to the Cooperative Patent Classification\footnote{\url{http://www.cooperativepatentclassification.org/index.html}}, and also implies that our research will not be updated and cannot make predictions specific to this system that can be tested in the future. More generally, we do recognize that nothing guarantees external validity; one could even argue that if the USPCS is discontinued and other classification systems are not, it shows that the USPCS has specificities and therefore it is not representative of other classification systems. Nevertheless, we think that the USPCS had a major influence on technology classifications and is the best case study to start with.

\subsection{The early history of the USPCS}

The U.S. patent system was established on 31st July 1790, but the need for examination was abolished 3 years later and reestablished only in 1836. As a result, there was no need to search for prior art and therefore the need for a classification was weak. 

The earliest known official subject matter classification appeared in 1823 as an appendix to the Secretary of State's report to the Congress for that year \citep{rotkin1999history}. It classified 635 patents models in 29 categories such as ``Bridges and Locks'', 1184 in a category named ``For various purposes'', and omitted those which were not ``deemed of sufficient importance to merit preservations''. 

In 1829, a report from the Superintendent proposed that with the prospect of the new, larger apartments for the Patent office, there would be enough room for a systematic arrangement and classification of models. He appended a list of 14 categories to the report.\footnote{The main titles were Agriculture, Factory machine, Navigation, Land works, Common trades, Wheel carriages, Hydraulicks (the spelling of which was changed in 1830), Calorific and steam apparatus, Mills, Lever and screw power, Arms, Mathematical instruments, Chemical compositions and Fine arts.}

In 1830 the House of representatives ordered the publication of a list of all patents, which appeared in December 1830/January 1831 with a table of contents organizing patents in 16 categories, which were almost identical to the 14 categories of 1829 plus ``Surgical instruments'' and ``Horology''.\footnote{An interesting remark on this classification \citep{rotkin1999history} is that it already contained classes based on industry categories (agriculture, navigation, \dots) and classes based on a ``specific mechanical force system'' (such as Lever and screw power).}

In July 1836, the requirement of novelty examination came into effect, making the search for prior art more pressing. Incidentally, in December the Patent office was completely destroyed by a fire. In 1837, a new classification system of 21 classes was published, including a Miscellaneous class and a few instances of cross noting\footnote{The first example given by \citet{rotkin1999history} is a patent for a pump classified in both ``Navigation'' and in ``Hydraulics and Hydrostatics''}. The following year another schedule was published, with some significant reorganization and a total number of classes of 22.

A new official classification appeared in 1868 and contained 36 main classes. Commenting on this increase in the number of classes, the Commissioner of patents wrote that \citep{rotkin1999history}
\begin{quote}
``The number of classes has risen from 22 to 36, a number of subjects being now recognized individually which were formally merged with others under a more generic title. Among these are builder's hardware, felting, illumination, paper, and sewing machines, to each of which subject so much attention has been directed by inventors that a division became a necessity to secure a proper apportionment of work among the corps of examiners.'' 
\end{quote}
Clearly, one of the rationale behind the creation and division of classes is to balance the class sizes, but this was not only to facilitate search. This class schedule was designed with administrative problems in mind, including the assignment of patent applications to the right examiners and the ``equitable apportionment of work among examiners'' \citep{rotkin1999history}.

Shortly after 1868 a parallel classification appeared, containing 176 classes used in the newly set up patent subscription service. This led to a new official classification containing 145 classes and published as a book in 1872. The number of classes grew to 158 in 1878 and 164 in 1880. \citet{rotkin1999history} note that the 1880 classification did not contain any form of cross-noting and cross references, by contrast to the 1872 classification. In 1882 classification reached 167 classes and introduced indentation of subclasses at more than one level. The classification of 1882 also introduced a class called ``Electricity'', long before this general purpose technology fully reached its potential.

In 1893 it was made clear in the annual report that a Classification division was required ``so that [the history of invention] would be readily accessible to searchers upon the novelty of any alleged invention''. After that, the need for a classification division (and the associated claim for extra budget) was consistently legitimated by this need to ``oppose the whole of prior art'' to every new application. In 1898 the ``Classification division'' was created with a head, two assistants and two clerks, with the purpose of establishing clearer classification principles and reclassifiying all existing patents. This marked the beginning of professional classification at the USPTO.

Since then the classification division has been very active and the patent classification system has evolved considerably, as we document extensively in this paper. But before, we need to explain the basic organizing principles of the classification system.

\subsection{Rationale and organization of the modern USPCS}
\label{section:USPCSrationale}
The USPCS attributes to each patent at least one subject matter. A subject matter includes a main class, delineating the main technology, and a subclass, delineating processes, structural features and functional features. All classes and most subclasses have a definition. Importantly, these are the patent claims which are classified, not the whole patent itself. The patent inherits the classification of its claims; its main classification is the classification of its main (``most comprehensive'') claim.

There are different types of patents, and they are translated into different types of classes.
According to the USPTO\footnote{\url{http://www.uspto.gov/web/offices/pac/mpep/s1502.html}}, ``in general terms, a utility patent protects the way an article is used and works, while a design patent protects the way an article looks.'' The ``classification of design patents is based on the concept of function or intended use of the industrial design disclosed and claimed in the Design patent.''\footnote{\url{http://www.uspto.gov/page/seven-classification-design-patents}}.

During the 19$^{th}$ century classification was based on which industry or profession was using the invention, for instance ``Bee culture'' (449) or ``Butchering'' (452). The example of choice \citep{falasco2002bases,uspto2005handbook,strumsky2012using} is that of cooling devices which were classified separately if they were used to cool different things, such as beer or milk. Today's system would classify both as cooling devices into the class ``Heat exchange'' (165), which is the utility or function of the invention. Another revealing example \citep{schmookler1966invention,griliches1990patent} is that a subclass dealing with the dispensing of liquids contains
both a patent for a water pistol and one for a holy water dispenser.
This change in the fundamental principles of classification took place at the turn of the century with the establishment of the Classification division \citep{falasco2002bases,rotkin1999history}. Progressively, the division undertook to redesign the classification system so that inventions would be classified according their utility. The fundamental principle which emerged is that of ``utility classification by \emph{proximate} function'' \citep{falasco2002bases} where the emphasis on ``proximate'' means that it is the fundamental function of the invention, not some example application in a particular device or industry. For instance ``Agitating'' (366) is the relevant class for inventions which perform agitation, whether this is to wash clothes, churn butter, or mix paint \citep{simmons2014categorizing}.
Another classification by utility is the classification by effect or product, where the result may be tangible (e.g. Semiconductors device and manufacture, 438) or intangible (e.g. Audio signal system, 381). Finally, the classification by structure (``arrangement of components'') is sometimes used for simple subject matter having general function. This rationale is the most often used for chemical compounds and stock material. It is rarely used for classes and more often used at the subclass level \citep{uspto2005handbook}

Even though the classification by utility is the dominant principle, the three classification rationales (by industry, utility and structure) coexist. Each class ``reflects the theories of classification that existed at the time it was reclassified'' \citep{uspto2005handbook}.
In addition, the system keeps evolving as classes (and even more so subclasses) are created, merged and split. New categories emerge when the need is felt by an examiner and approved by the appropriate Technology Center; in this case the USPCS is revised through a ``Classification order'' and all patents that need to are reclassified \citep{strumsky2012using}. An example of how subclasses are created is through alpha subclasses. Alpha subclasses were originally informal collections created by patent examiners themselves to help their work, but were later incorporated into the USPC. They are now created and used as temporary subclasses until they become formalized \citep{falasco2002united,uspto2005handbook}. When a classification project is completed, a classification order is issued, summarising the changes officially, and all patents that need to are, in principle, reclassified.

One of the latest class to have been created is ``Nanotechnology (977)'', in October 2004. As noted by \citet{strumsky2012using}, using the current classification system one finds that after reclassification the first nanotechnology patent was granted much earlier\footnote{1986 for \citet{strumsky2012using}, 1978 for \citet{paradise2012claiming} and 1975 according to \citet{strumsky2015identifying} and to the data that we use here (US3896814). Again, these differences reflect the importance of reclassification.}. According to \citet{paradise2012claiming}, large federal research funding led to the emergence of ``nanotechnology'' as a unifying term, which became reflected in scientific publications and patents. Because nanotechnologies were new, received lots of applications and require interdisciplinary knowledge, it was difficult to ensure that prior art was reviewed properly. The USPTO engaged in a classification project in 2001, which started by defining nanotechnologies and establishing their scope, through an internal process as well as by engaging with other stakeholders such as users or other patent offices. In 2004 the Nanotechnology cross-reference digest was established; cross-reference means that this class cannot be used as a primary class. \citet{paradise2012claiming} argues that class 977 has been defined with a too low threshold of 1 to 100 nanometers. Also, reclassification has been encouraged but is not systematic, so that many important nanopatents granted before 2004 may not be classified as such.

Another example of class creation worth mentioning is given by \citet{erdi2013prediction} who argue that the creation of ``Fabric (woven, knitted, or nonwoven textile or cloth, etc.)'' (442) created in 1997, could have been predicted based on clustering analysis of citations. \citet{kyebambe2017forecasting} recently generalized this approach, by formulating it as a classical machine learning classification problem: patent clusters are characterized by sets of features (citations, claims, etc.), and only some patent clusters are later on recognized as ``emerging technology'' by being reclassified into a new USPCS main class. In this sense, USPCS experts are labelling data, and \citet{kyebambe2017forecasting} developed a method to create clusters and train machine learning algorithms on the data labelled by USPCS experts.

Finally, a last example is that of organic chemistry\footnote{see \url{http://www.uspto.gov/page/addendum-reclassification-classes-518-585}}. Class 260 used to contain the largest array of patent documents but it was decided that this class needed to be reclassified ``because its concepts did not necessarily address new technology and several of its subclasses were too difficult to search because of their size.''. To make smaller reclassification projects immediately available it was decided to split the large class into many individual classes in the range of Classes 518-585. Each of these classes is ``considered an independent class under the Class 260 umbrella''; many of these classes have the same general name such as ``Organic coumpounds \--- part of the class 532-570 series''\footnote{These classes also have a hierarchy indicated by their number, as subclasses within a class schedule usually do.}

As argued by \citet{strumsky2012using}, this procedure of introducing new codes and modifying existing ones ensures that the current classification of patents is consistent and makes it possible to study the development of technologies over a long period of time. However, while looking at the past with today's glasses ensures that we look at different periods of the past in a consistent way, it is not the same as reporting what the past was in the eyes of those who lived it. In this sense, we believe that it is also interesting to try and reconstruct the classification systems that were in place in the past. We now describe our preliminary attempt to do so, by listing available sources and constructing a simple count of the number of classes used in the past.

\subsection{Dataset construction}
\label{section:data-construction}

Before describing the data construction in details, let us state clearly three important caveats.

First, we focus on main classes, due to the difficulty of collecting historical data at the subclass level. This is an important omission and avenue for further research. Investigating the complete hierarchy could add significant insight, for instance by contrasting ``vertical'' and ``horizontal'' growth of the classification tree, or by exploiting the fact that different layers of system play a different role for search \citep{uspto2005handbook}.

Second, we limit our investigations to Primary (``OR'') classes, essentially for simplicity. Multiple classifications are indeed very interesting and would therefore warrant a complete independent study. Clearly, the fact that multiple classifications can be used is a fundamental feature of the current USPCS. In fact it is a key feature of its evolution: as noted above ``cross-noting'' was common in some periods and absent in others, and a recent example of a novel class \--- Nanotechnology \-- happens to be an XR-only class (i.e., used only as secondary classification). Here we have chosen to use only OR classes because it allows us to show the main patterns in relatively simple way. Of course some of our results, in particular those of Section \ref{section:reclass}, are affected by this choice, and further research will be necessary to evaluate the robustness of our results. That said, OR classifications, which are used on patent applications to find the most appropriate examining division \citep{falasco2002united}, are arguably the most important.

Third, we limit our investigation to the USPCS, as justified in the beginning of Section \ref{section:data}. We have good reasons for choosing the USPCS in this study, which aims at giving a long-run picture. However, for studying the details of reclassification patterns and firmly establishing reclassification and classification system changes as novel and useful indicators of technological change, future research will need to establish similar patterns in the IPC or CPC.

As a result of these choices, our aim is to build a database\footnote{Our data is available at \url{https://dataverse.harvard.edu/dataset.xhtml?persistentId=doi:10.7910/DVN/ZJCDCE}} of 1) the evolution of the USPCS primary classes, and 2) the reclassification of patents from one class to the other.  To do this we relied on several sources. 

First, our most original data collection effort concerns the historical number of classes. For the early years our main sources are \citet{bailey1946history} and \citet{rotkin1999history}, complemented by \citet{reingold1960us} and the ``Manual of Classification'' for the 5 years within the period 1908\---1923.
For the 1950\---60's, we used mostly a year-specific source named ``General information concerning Patents'' which contained a sentence like ``Patents are classified into $x$ classes''. Unfortunately, starting in 1969 the sentence becomes ``Patents are classified into more than 310 classes''.
We therefore switched to another source named ``Index of patents issued from the United States Patent Office'', which contains the list of classes. Starting 1963, it contains the list of classes with their name and number on a separate page\footnote{We had to make some assumptions. In the 1960's, Designs appeared subdivided into ``Industrial arts'' and ``Household, personal and fine arts'', so we assumed that the number of design classes is 2, up to the year 1977 where Design classes appear with their name and number. We implicitly assume that prior to 1977 the design classes were actually subclasses, since in 1977, there were 39 Design classes, whereas the number of (sub)classes used for design patents in 1976 was more than 60. It should be noted though that according to the dates established, some of the current design classes were created in the late 60's. Another issue was that for 1976 the number of Organic compound classes was not clear \-- we assumed it was 6, as listed in 1977. Finally, we sometimes had two slightly different values for the same year due to contradictory sources or because the sources refer to a different month.\label{footnoteDesign}}.
For 1985, we used a report of the Office of Technology Assessment and Forecast (OTAF) of the Patent and Trademark Office \citep{otaf1985}. 
For the years 2001 to 2013, we collected data from the Internet Archive.\footnote{\url{https://archive.org/index.php} where we can find the evolution of the url \url{http://www.uspto.gov/web/patents/classification/selectnumwithtitle.htm}. We added the class ``001'' to the count.}
As of February 2016 there are 440 utility classes (including the Miscellaneous 001 and and the ``Information storage'' G9B (established in 2008)), 33 design classes, and the class PLT ``Plant", giving a total of 474 classes.\footnote{
The list of classes available with their dates established contains 476 classes, but it does not contain 001, and it contains 364, 389, and 395 which have been abolished. We removed the abolished classes, and for Figs \ref{fig:Nclasses} and  \ref{fig:Heaps} we assumed 001 was established in 1899.}. 

Second, to obtain reclassification data we matched several files. We obtained ``current'' classifications from the Master Classification File (version mcfpat1506) for patents granted up to the end of June 2015. We matched this with the Patent Grant Authority File (version 20160130) to obtain grant years\footnote{
We first removed 303 patents with no main (OR) classification, and then 92 patents dated January 1st 1800. We kept all patent kinds.}.
To obtain the classification at birth, we used  the file ``Patent Grant Bibliographic (Front Page) Text Data (January 1976 -- December 2015)'', provided by the USPTO\footnote{at \url{https://bulkdata.uspto.gov/} (Access date: January 7, 2018)}, from which we also gathered citation data.

\section{Dynamics of the number of classes and Heaps' law}
\label{section:Nclass}

\begin{figure*}[!ht]
	\centering
		\includegraphics[scale=0.75]{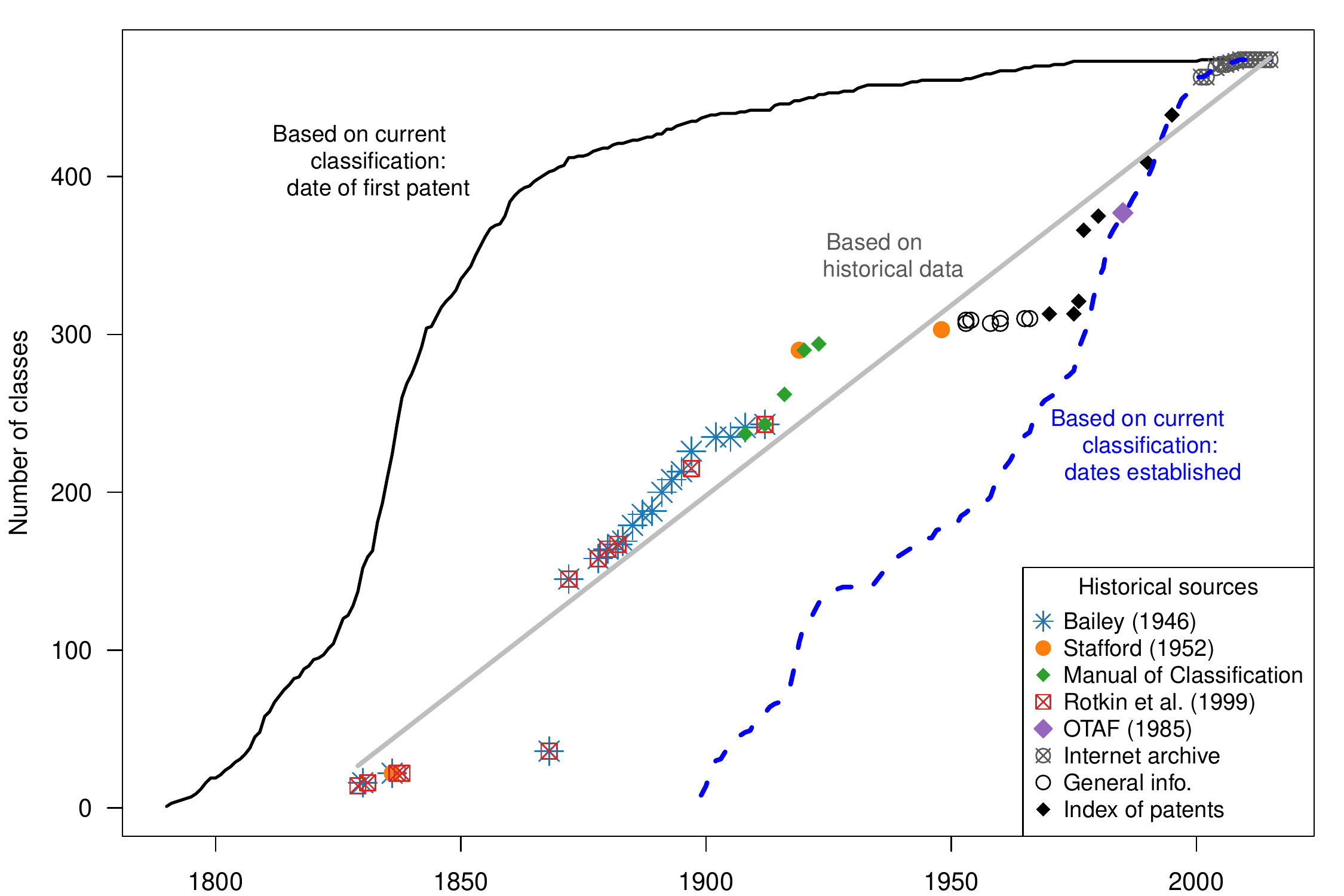}
	\caption{Evolution of the number of distinct classes.}
	\label{fig:Nclasses}
\vspace{5mm}
\includegraphics[scale=0.75]{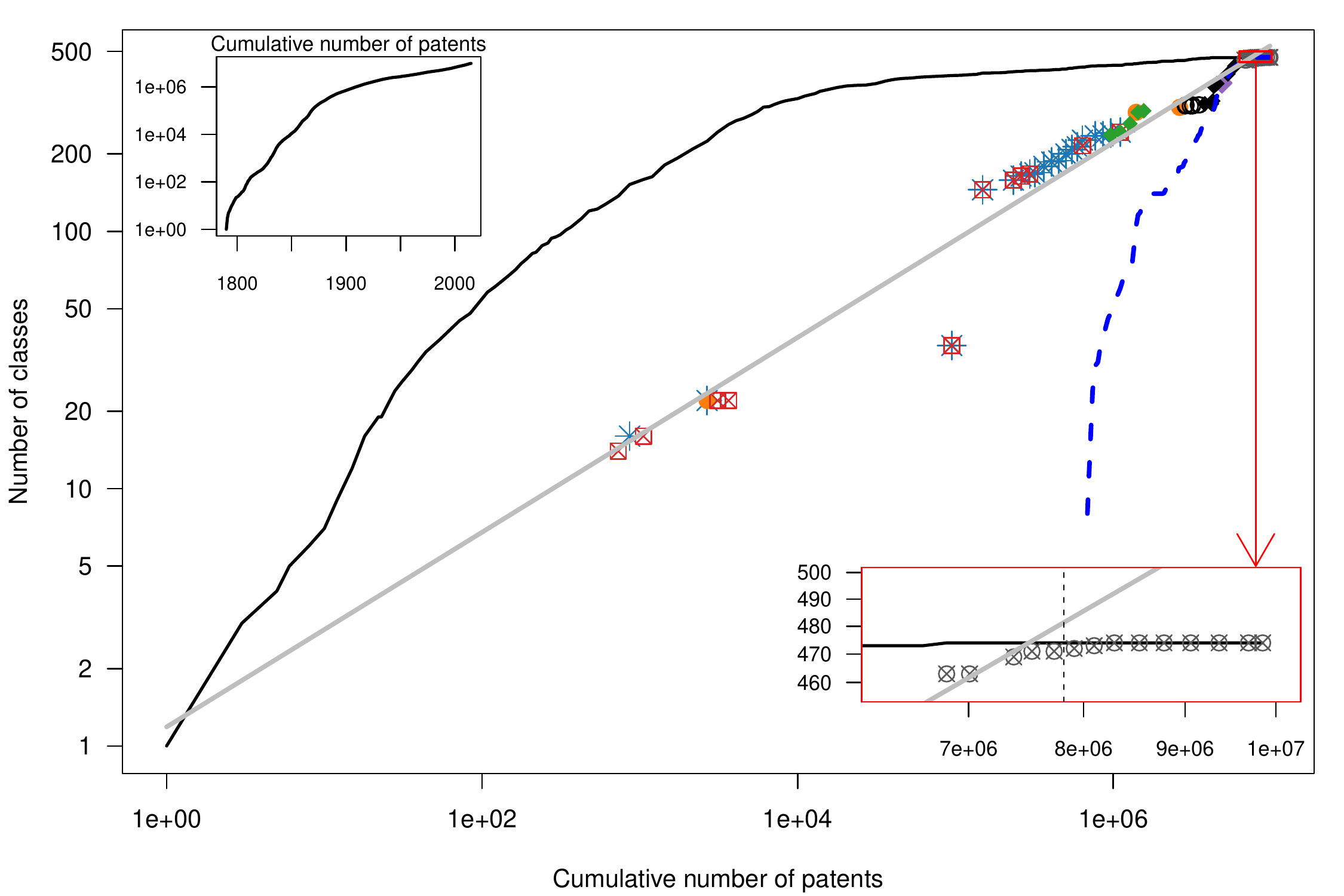}
\caption{Heaps' law.}
\label{fig:Heaps}
\end{figure*}

Our first result concerns the growth of the number of classes (Fig. \ref{fig:Nclasses}), which we have computed using three different methods. 

First, we used the raw data collected from the historical sources mentioned in Section \ref{section:data-construction}. 
Quite unexpectedly, the data suggests a linear growth, with appreciable fluctuations mainly due to the introduction of an entirely new system in 1872 and to design classes in 1977 (see footnote \footref{footnoteDesign}). The grey line shows the linear fit with an estimated slope of 2.41 (s.e. 0.06) and $R^2$ of 0.96 (we treat years with no data as NA, but filling them with the figure from the last observed year does not dramatically affect the results).
 
Second, we have computed, using the Master Classification File for June 2015, the number of distinct classes in which the patents granted up to year $t$ are classified (black line). To do so, we have used all classes in which patents are classified (i.e. including cross-reference classes).\footnote{The (reconstructed) number of classes is slightly lower if we consider only Primary classes, because some classes are used only as a cross-reference, never as primary class. These classes are
    902: Electronic funds transfer,
    903: Hybrid electric vehicles,
    901: Robots,
    930: Peptide or protein sequence,
    977: Nanotechnology,
    976: Nuclear technology,
    968: Horology,
    987: Organic compounds containing a bi, sb, as, or p atom or containing a metal atom of the 6th to 8th group of the periodic system,
    984: Musical instruments,
    G9B: Information storage based on relative movement between record carrier and transducer.
} The pattern of growth is quite different from the historical data. If we consider only the post-1836 data, the growth of the number of classes is sublinear \--- less and less classes are introduced every year. Before 1836, the trend was linear or perhaps exponential, giving a somewhat asymmetric S-shape to the overall picture. 

Third, we computed the growth of the number of classes based on the dates at which all current classes were established (blue line)\footnote{Collected from \url{https://www.uspto.gov}, page USPCS dates-established}. According to this measure, the first class was created in 1899, when the reorganization of classification started with the creation of the classification division\footnote{``Buckles, Buttons, clasps, etc.'' is an example of a class that was created early under a slightly different name (1872 according to \citet{simmons2014categorizing}, see \citet{bailey1946history} for details) but has a posterior ``date established'' (1904 according to the USPTO). Another example is ``Butchering''.}. 

Fig. \ref{fig:Heaps} displays the number of classes against the number of patents in a log-log scale. In many systems, it has been found that the number of categories grows as a power law of the number of items that they classify, a result known as Heaps' law (for an example based on a classification system \---the medical subject headings\--- instead of a language, see \citet{petersen2016triple}). Here we find that using the 2015 classification, Heaps' law is clearly violated\footnote{It is possible to obtain a good fit by limiting the fit to the latest periods, however this is arbitrary, and gives a very low Heaps' exponent, leaving unexplained the creation of the vast majority of classes.}. Using the historical data, Heaps' law appears as a reasonable approximation. We estimate the Heaps' exponent to be $0.378$ with standard error of 0.010 and $R^2=0.95$.
The inset on the bottom right of Fig. \ref{fig:Heaps} shows that for the latest years, Heaps' law fails: for the latest 2 million patents (about 20\% of the total), almost no classes were created. We do not know whether this slowdown in the introduction of classes is due to a slowdown of radical innovation, or to a more institutionally-driven reason such as a lack of investment in the USPCS due to the expected switch to the Cooperative Patent Classification. Since the joint classification system was first announced on 25 October 2010 \citep{blackman2011classification}, we show this date (more precisely, patent number 7818817 issued on the $26^{th}$) as a suggestive indicator (dashed line on the inset). Another consideration is that the system may be growing more ``vertically'', in terms of the number of layers of subclasses \--- unfortunately here we have to focus on classes, so we are not able to test for this.

\section{The size distribution and the age-size relationship}
\label{section:sizedist}

Besides the creation and reorganization of technological categories, we are interested in their growth and relative sizes. More generally, our work is motivated by the Schumpeterian idea that the economy is constantly reshaping itself by introducing novelty \citep{dopfer2004micro,saviotti2004economic}. 
The growth of technological domains has been deeply scrutinized in the economics of technical change and development \citep{schumpeter1934theory, dosi1982technological, pasinetti1983structural, pavitt1984sectoral, freeman1997economics, saviotti1996technological,malerba2002sectoral}. A recurring theme in this literature is the high heterogeneity among sectors. When sectors or technological domains grow at different rates, structural change occurs: the relative sizes of different domains is modified. To study this question in a parsimonious way, one may opt for a mesoscale approach, that is, study the size distribution of categories. 

Our work here is most directly related to \citet{carnabuci2013distribution} who first showed on data for 1963\---1999 that the size distribution of classes is close to exponential. This is an interesting and at first surprising finding, because based on the assumption that all domains grow at the same average rate stochastic growth models such as \citet{gibrat1931inegalites} or \citet{yule1925mathematical} predict a Log-normal or a Pareto distribution, which are much more fat tailed. Instead, we do not see the emergence of relatively very large domains, and this may at first suggest that older sectors do not keep growing as fast as younger ones, perhaps due to technology life-cycles \citep{vernon1966international, klepper1997industry,andersen1999hunt}. However, as we will discuss, we are able to explain the exponential size distribution by keeping Gibrat's law, but assuming that categories are split randomly.

\subsection{The size distribution of categories}

\begin{figure}[H]
\centering
\includegraphics[scale=0.6]{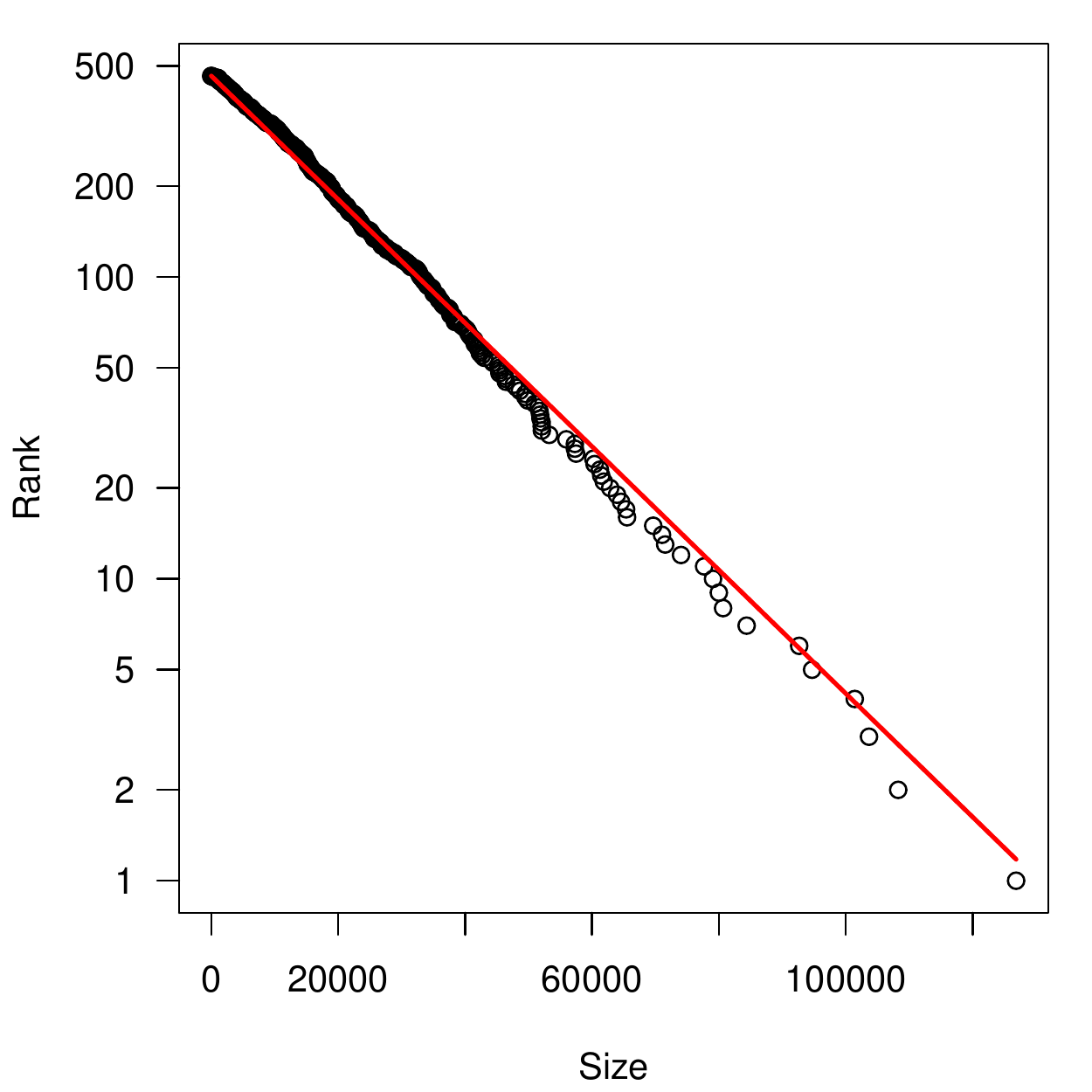}		
\caption{Rank-size relationship.}
\label{fig:ranksize}
\end{figure}

In this section we study the size distribution of classes, where size is the number of patents in 2015 and classes are defined using the current classification system. We use only the primary classification, so we have only 464 classes.
Fig. \ref{fig:ranksize} suggests a linear relationship between the size of a class and the log of its rank, that is, class sizes are exponentially distributed\footnote{
For simplicity we used the (continuous) exponential distribution instead of the more appropriate (discrete) geometric distribution, but this makes no difference to our point. We have not rigorously tested whether or not the exponential hypothesis can be rejected, because the proper hypothesis is geometric and classical test statistics such as Kolmogorov-Smirnov do not easily apply to discrete distributions. Likelihood ratio tests interpreted at the 5\% level showed that it is possible to obtain better fits using two-parameters distributions that extends the exponential/geometric, namely the Weibull and the Negative binomial, especially after removing the two smallest categories which are outliers (contain 4 and 6 patents) and are part of larger series (532 and 520).}. To see this, let $p(k)$ be the probability density of the sizes $k$. If it is exponential, it is $p(k)=\lambda e^{-\lambda k}$. By definition, the rank $r(k)$ of a class of size $k$ is the number of classes that have a larger size, which is $r(k)=N \int_{k}^{\infty} \lambda e^{-\lambda x} dx = N e^{-\lambda k}$,
where $N$ is the number of classes. This is equivalent to size being linear in the logarithm of the rank. We estimate the parameter $\lambda$ by maximum likelihood and obtained $\hat{\lambda}=4.71 \times 10^{-5}$ with standard error $0.22 \times 10^{-5}$. Note that $\hat{\lambda}$ is one over the mean size, 21223. We use this estimate to plot the resulting fit in Fig. \ref{fig:ranksize}.

It is interesting to find an exponential distribution, since one may have expected a power law, which is quite common as a size distribution, and appears often with Heaps' law \citep{lu2010zipf,petersen2016triple}. 
Since the exponential distribution is a good representation of the data, it is worth looking for a simple mechanism that generates this distribution, which we will do in Section \ref{section:model}. But since many models can generate an exponential distribution we first need to present additional empirical evidence that will allow us to discriminate between different candidate models.

\subsection{The age-size relationship}

To determine whether older classes contain more patents than younger ones, we first need to note that there are two ways of measuring age: the official date at which the class was established, and the year in which its first patent was granted. As expected, it appears that the year in which a class is established is always posterior to the date of its first patent\footnote{
Apart from class 532. We confirmed this by manually searching the USPTO website. 532 is part of the Organic compound classes, which have been reorganized heavily, as discussed in Section \ref{section:USPCSrationale}
}.

\begin{figure}[H]
\centering
\includegraphics[scale=0.55]{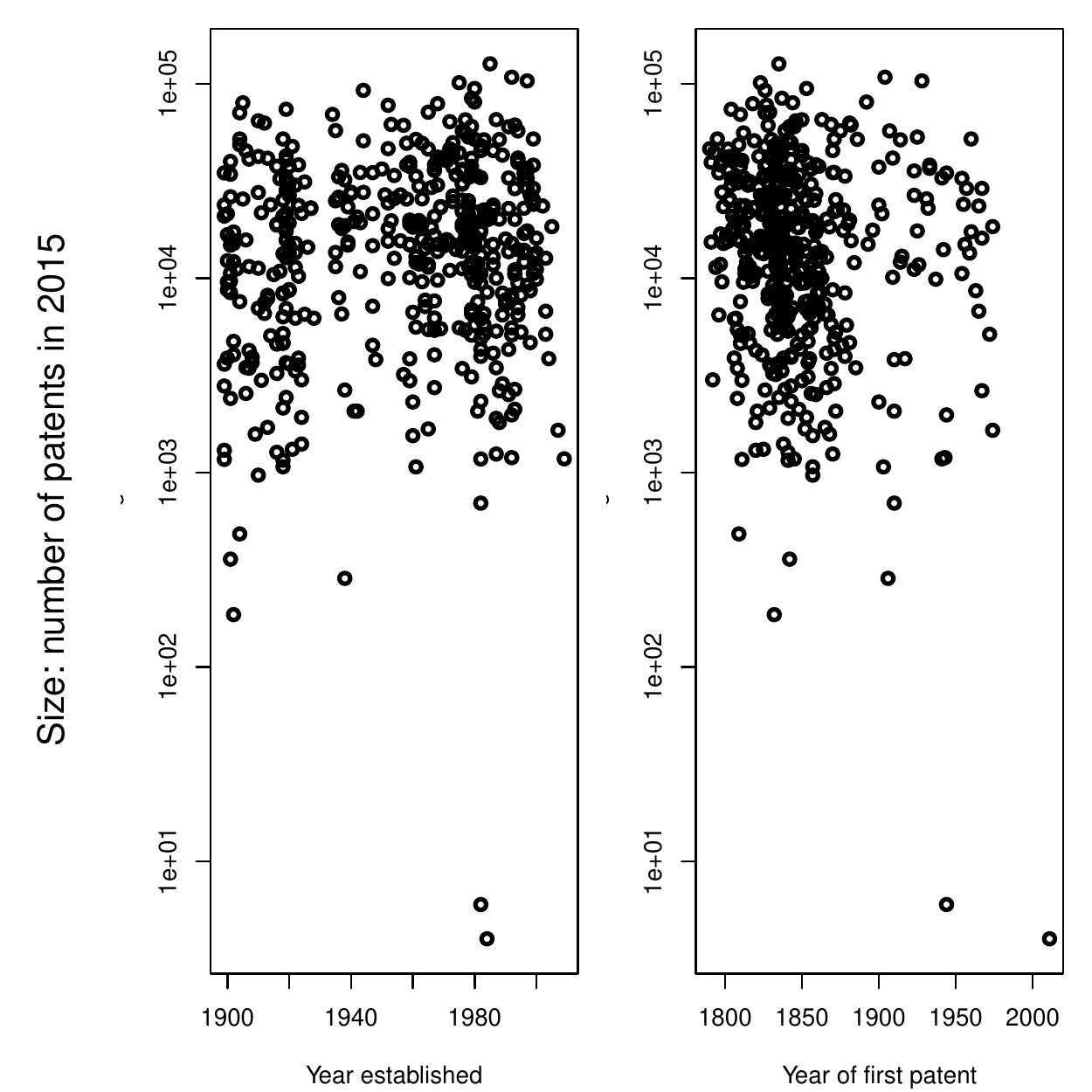}
\caption{Age-size relationship.}
\label{fig:agesize}
\end{figure}

Since these two ways of measuring age can be quite different, we show the age-size (or rather size-birth date) relationship for both in Fig. \ref{fig:agesize}. If stochastic growth models without reclassification were valid, we would observe a negative slope, that is, newer classes should have fewer patents because they have had less time for accumulation from random growth.
Instead, we find no clear relationship. In the case of the year established, linear regressions indicated a positive relationship significant at the 10\% but not at the 5\% confidence level, whether or not the two ``outliers'' were removed. Using a log-linear model, we found a significant coefficient of 0.004 after removing the two outliers. In the case of the year of the first patent, the linear model indicated no significant relationship, but the log-linear model delivered a highly significant negative coefficient of -0.005 (which halves and becomes significant at the 10\% level only once the two outliers are removed); In all 8 cases (two different age variables and two different models, removing outliers or not) the $R^2$ was between 0.001 and 0.029. 

We conclude that these relationships are at best very weak, and in one case of the ``wrong'' sign (with classes established in recent years being on average larger). Whether they are significant or not, our point here is that their magnitude and the goodness of fits are much lower than what one would expect from growth-only models such as \citet{simon1955class}, or its modification with uniform attachment (to match the exponential size distribution). We will come back to the discussion of models later, but first we want to show another empirical pattern and explain why we think reclassification and classification system changes are interesting indicators of technological change.

\section{Reclassification activity as an indicator of technological change}
\label{section:reclass}

It seems almost tautological to say that a radical innovation is hard to categorize when it appears. If an innovation is truly ``radical'', it should profoundly change how we think about a technology, a technological domain, or a set of functions performed by technologies. If this is the case a patent related to a radical innovation is originally hard to classify. It is likely that it will have to be reclassified in the future, when a more appropriate set of concepts has been developed and institutionalized (that is, when the community of technologists have codified a novel understanding about the radical innovation). It is also well accepted that radical innovations may create a new wave of additional innovations, which may or may not cluster in time \citep{silverberg2003breaking} but when they are general purpose we do expect a rise in innovative activity \citep{bresnahan1995general}. A less commented consequence of the emergence and diffusion of General Purpose Technologies (GPTs) is that both due to the sheer increase in the number of patents in this technology, and to the impact of this technology on others, we should expect higher classification volatility. Classification volatility is to be expected particularly in relation to GPTs because by definition GPTs interact with existing technologies and create or reorganize interactions among existing technologies. From the point of view of the classification, the very definition of the objects and their boundaries are transformed. In short, some categories become too large and need to be split; some definitions become obsolete and need to be changed; and the ``best'' grouping of technologies is affected by the birth and death of conceptual relationships between the function, industry of origin or application, and structural features of technologies.

In this section we provide a preliminary study. First we establish that this indicator does exist (reclassification rates can be quite high, reaching 100\% if we look far enough in the past). Second, we show that reclassified patents are more cited. Third, we show that reclassification can take place across fairly distant technological domains, as measured by 1-digit NBER categories. Fourth, we discuss three examples of novel classes.

\subsection{Reclassification rates}

How many patents have been reclassified? To start with, since no classification existed prior to 1829, all patents published before that have been ``(re)classified'' in the sense that their category has been determined several and potentially many years after being granted. The same applies to all patents granted at times where completely different classification systems prevailed, which is the case before 1899. In modern times, classification has evolved but as discussed in Section \ref{section:data}, the overall classification framework put in place at the turn of the century stayed more or less the same. For the period after 1976, we know the original classification of each patent because we can read it on the digitized version of the original paper (see Section \ref{section:data-construction}). After extensive efforts in parsing the data and a few manual corrections, we found an original class for 99.45\% of the post-1976 patents in the Master Classification File mcfpat1506. Out of these 5,615,525 patents, 412,724 (7.35\%) have been reclassified. There are 789 distinct original classes, including 109 with only 1 patent (apart from data errors, this can come from original classes that had no post-1976 patents classified in them). All current classes have been used as original classes except ``001'' which is only used as a miscellaneous class in which they are reclassified\footnote{We removed US6481014.}.

\begin{figure}[ht]
\centering
\includegraphics[scale=0.75]{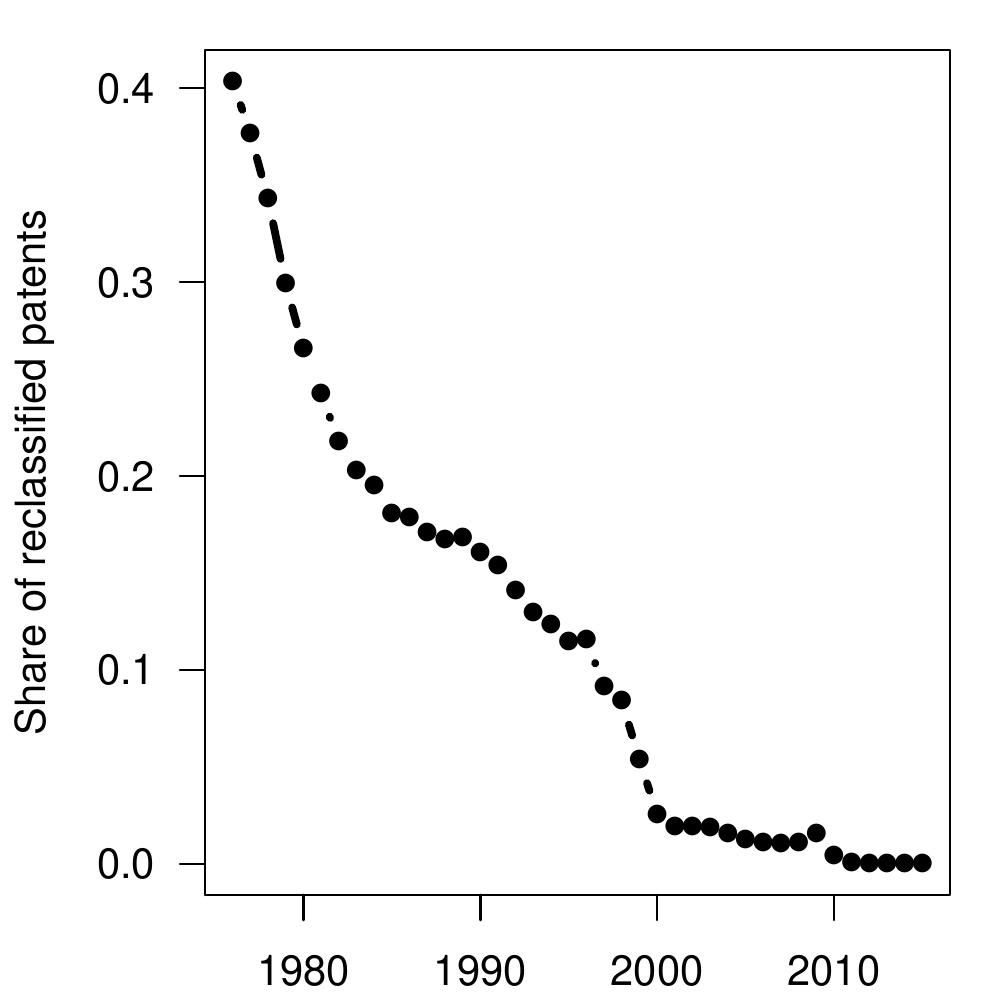}
\caption{Share of patents granted in a given year that are in a different class in 2015, as compared to when they were granted.}
\label{fig:sharereclass}
\end{figure}

Figure \ref{fig:sharereclass} shows the evolution of the reclassification rate, defined as the share of patents granted in year $t$ which have a different classification in 2015 than in $t$. It appears that as much as 40\% of the 1976's patents belong to a different class now than when they first appear. This reclassification rate declines sharply after that, reaching about 10\% in the 1990's and almost zero thereafter. This is an expected result, since the longer the time since granting the patent, the higher the chances that the classification system has changed.

\subsection{Are reclassified patents more cited?}

Since there is an established relationship between patent value and the number of citations received \citep{hall2005market}, it is interesting to check if reclassified patents are more cited. Of course, we are only observing correlations, and the relationship between citations and reclassification can work in multiple ways. A plausible hypothesis is that the more active is a technological domain (in terms of new patents and thus new citations being made), the more likely it is that there will be a need for reclassification, if only to keep the classes at a manageable size\footnote{Relatedly, as noted by a referee, if patent examiners are also responsible for reclassification, then their prior art search might be oriented towards patents that they have re-classified, for which their memory is more vivid.}. Another hypothesis is that highly innovative patents are intrinsically ambiguously defined in terms of the classification system existing when they first appear. In any case, since we only have the class number at birth and the class number in 2015, we cannot make subtle distinctions between different mechanisms. However, we can check whether reclassified patents are on average more cited, and we can do so after controlling for the grant year and class at birth. 

Table \ref{table:cit} shows basic statistics\footnote{We count citations made to patents for which we have reclassification data, from patents granted until June 2015. We removed duplicated citations}. Reclassified patents constitute 7.35\% of the sample, and have received on average more than 24 citations, which is more than twice as much as the non reclassified patents.

% latex table generated in R 3.4.2 by xtable 1.8-2 package
% Thu May 31 09:13:23 2018
\begin{table}[ht]
\centering
\begin{tabular}{|c|rrrr|}
  \hline
 & share & mean & median & s.d. \\ 
  \hline
All & 100.00 & 11.30 & 4.00 & 26.64 \\ 
  Non reclassified & 92.65 & 10.27 & 4.00 & 23.94 \\ 
  Reclassified & 7.35 & 24.29 & 11.00 & 47.40 \\ 
   \hline
\end{tabular}
\caption{Patent citations summary statistics.}
\label{table:cit}
\end{table}

\begin{figure}[ht]
\centering
\includegraphics[scale=0.6]{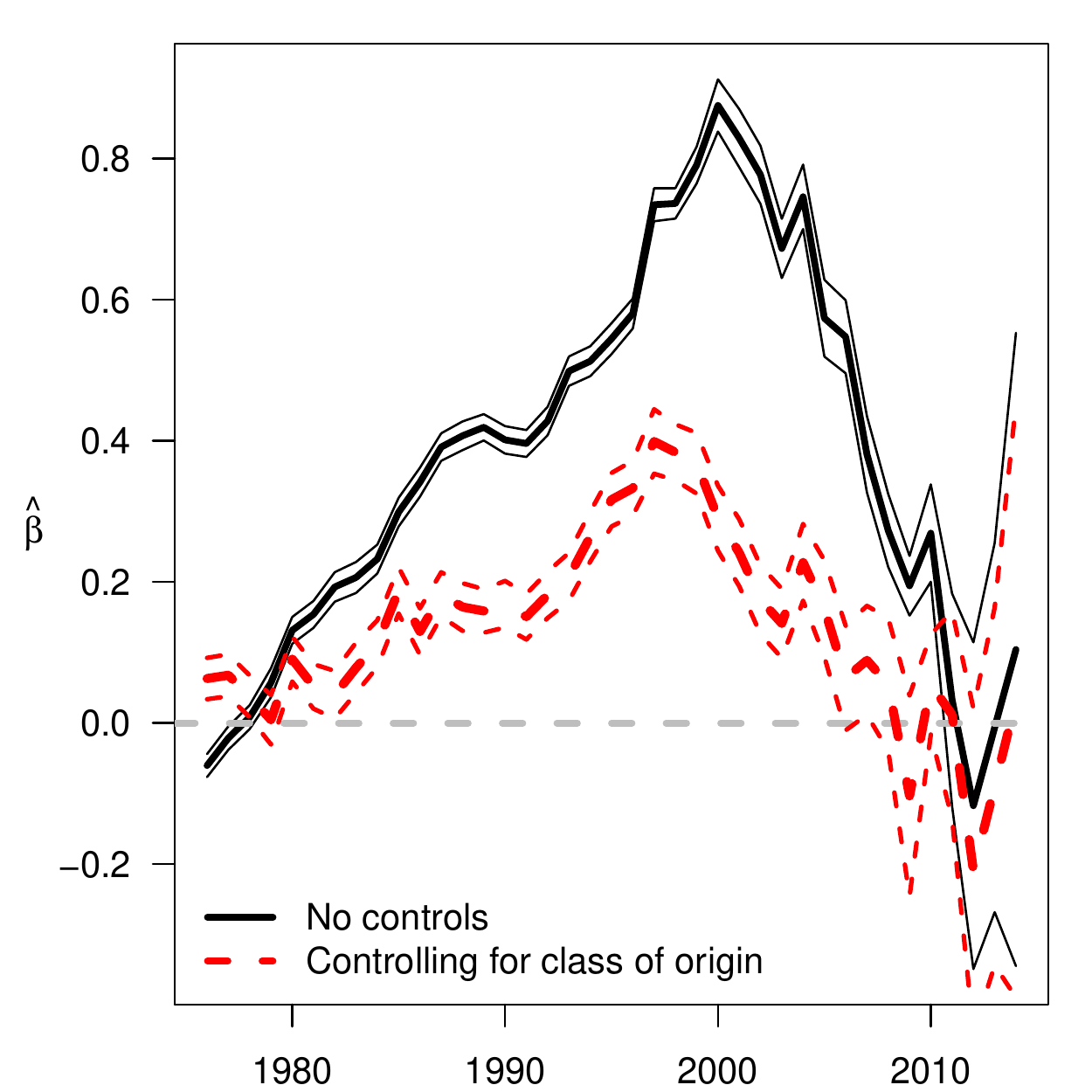}
\caption{Coefficient of the year-specific regressions of the log of citations received on the reclassification dummy (including dummies for the class of origin or not).}
\label{fig:reg-coeff-time-evol}
\end{figure}

We expect this result to be largely driven by the fact that older patents have both a higher chance to have been reclassified and a higher chance to have accumulated many citations. To investigate the relationship between reclassification and citations in more detail, we regressed the log of total citations received in 2015 on the reclassification dummy and on dummies for the class at birth, for each year separately (and keeping only the patents with at least one citation received, 76.6\%):
\[
\log(c_{i})=\alpha_t + \beta_t R_i + \sum_{j=1}^{J_t-1} \gamma_{j,t} D_{i,j}
\]
where $c_{i}$ is the number of citations received by patent $i$ between its birth (time $t$) and (June) 2015, $R_i$ is a dummy that takes the value of 1 if patent $i$ has a main class code in 2015 different from the one it had when it appeared (i.e. in year $t$), $J_t$ is the number of distinct classes in which the patents born in year $t$ were classified at birth, and $D_{i,j}$ is a dummy that takes the value of 1 if patent $i$ was classified in class $j$ at birth.

Note that we estimate this equation separately for every grant year. We include the class at birth dummies because this allows us to consider patents that are ``identical twins'' in the sense of being born in the same class in the same year. The coefficient $\beta$ then shows if reclassified patents have on average received more citations. The results are reported in Fig. \ref{fig:reg-coeff-time-evol}, showing good evidence that reclassification is associated with more citations received. As expected, recent years\footnote{2015 is excluded because no patents had been reclassified} are not significant since there has not been enough time for reclassification to take place and citations to accumulate (the bands represent standard approximate 95\% confidence intervals). We also note that controlling for the class at birth generally weakens the effect (red dashed line compared to black solid line).

\subsection{Reclassification flows}

\begin{figure*}[ht]
\centering
\includegraphics[width=\textwidth]{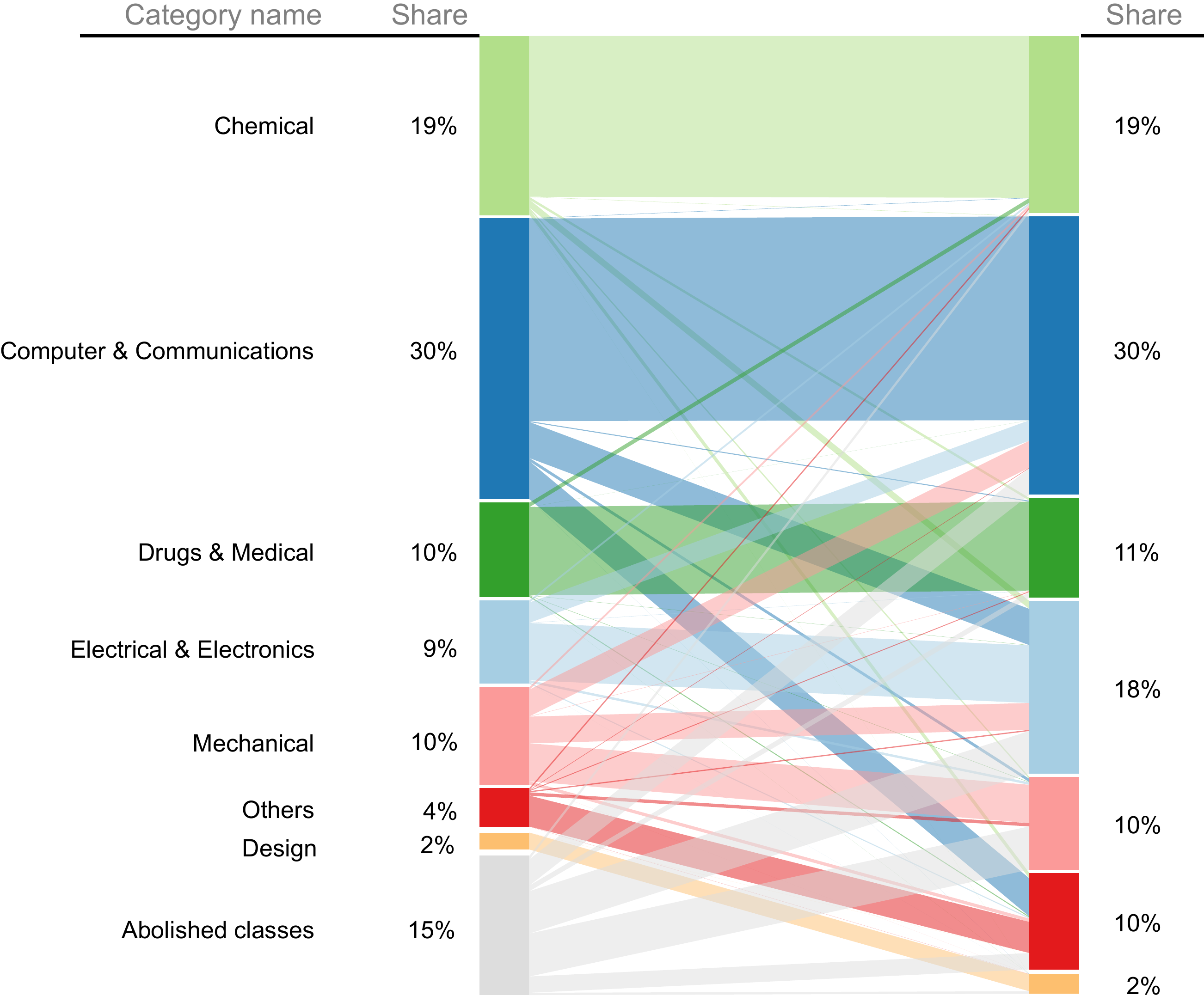}
\caption{Reclassification flows.}
\label{fig:reclassification_flow}
\end{figure*}

To visualize the reclassification flows, we consider only the patents that have been reclassified. As in \citet{wang2016technological} we want to construct a bipartite graph showing the original class on one side and the current class on the other side. Since we identify classes by their code number, a potentially serious problem may arise if classes are renumbered, although we believe this tends to be rare given the limited time span 1976\---2015. An example of this is ``Bee culture'' which was class number 6, but since 1988 is class number 449 and class number 6 does no longer exists. However, even in this case, even though these two classes have the same name, we do not know if they are meant to encompass the same technological domain and have just been ``renumbered'', or if other considerations prevailed and renumbering coincides with a more substantive reorganisation. An interesting extension of our work would be to use natural language processing techniques on class definitions to define a measure of reclassification distance more precisely and exclude mere renumbering.

To make the flow diagram readable and easier to interpret, we aggregate by using the NBER categories\footnote{For more details on the NBER categories, see the historical reference \citep{hall2001nber} and the recent effort by \citet{marco2015uspto} to attribute NBER (sub) categories to patent applications.}. To assign each class to a NBER category, we used the 2006 version of the NBER classification, which we modified slightly by classifying the Design classes separately, and classifying USPCS 850 (Scanning probe techniques and apparatus) in NBER 4 (Electrical) and USPCS PLT (Plant) in NBER 6 (Others). 

Fig. \ref{fig:reclassification_flow} shows the results\footnote{See the online version at \url{http://danielykim.me/visualizations/PatentReclassificationHJTcategory/}}. The share of a category means the fraction of reclassified patents whose primary class is in a particular NBER category. The width of the lines between an original category $i$ and a current category $j$ is proportional to the number of reclassified patents whose original class is in category $i$ and current class is in category $j$. Line colors indicate the original category.

We can see that patents originally classified in the categories Chemical tend to be reclassified in another class of the category Chemical. The same pattern is observed for the category Drugs. By contrast, the categories Computers \& Communications and Electrical \& Electronics display more cross-reclassifications, in line with \possessivecite{wang2016technological} findings on a restricted dataset. This may indicate that the NBER categories related to computers and electronics are not as crisply defined as those related to Chemical and Drugs, and may be suggestive of the general purpose nature of computers. This could also suggest that that these domains were going through a lot of upheaval during this time period. While there is some ambiguity in interpreting these patterns, they are not \emph{a priori} obvious and point to the same phenomenon as the correlation between citations and reclassifications: dynamic, impact-full, really novel, general purpose fields are associated to more taxonomic volatility.

\subsection{Three examples of novel classes}

We now complement the study by providing three examples of novel classes, chosen among recently created classes (and excluding cross-reference only classes). We proceed by looking at the origin of patents reclassified in the new class when it is created. We approximate this by looking at the patents that have been granted on a year preceding the birth year of a class, and now appear as reclassified into it. Note that we can determine the class of origin only for patents granted after 1976. We also give as example the oldest reclassified (utility) patent we can find. We discuss each class separately (see Table \ref{table:reclass_main} for basic statistics on each of the three example classes, and Table \ref{table:reclass_origin} for the source classes in each case (``Date'' is the date at which an ``origin'' class was established.)

% latex table generated in R 3.4.2 by xtable 1.8-2 package
% Thu Jan 11 14:28:31 2018
\begin{table}[ht]
\centering
\begin{tabular}{|p{15mm}p{20mm}p{15mm}p{15mm}|}
  \hline
Class Number & Date

 established & Size &  Size post 1976 \\ 
  \hline
442 & 1997 & 6240 & 2654 \\ 
  506 & 2007 & 1090 & 1089 \\ 
  706 & 1998 & 1270 & 1217 \\ 
   \hline
\end{tabular}
\caption{Basic information for the three novel classes described in the main text. Size is the number of patents that are classified in a class now but were granted before the class was created. Size post 1976 is the same, but excluding all pre-1976 patents, to be compared with the size of classes of origin in Table \ref{table:reclass_origin}.}
\label{table:reclass_main}
\end{table}

% latex table generated in R 3.4.2 by xtable 1.8-2 package
% Wed Jan 10 19:12:10 2018
\begin{table}[ht]
\centering
\begin{tabular}{|rp{45mm}cc|}
  \hline
 \multicolumn{4}{|c|}{Classes of origins for Class 442}\\
  \hline
Size & Title &  Num. & Date  \\ 
  \hline
2615 & Stock material or miscellaneous articles & 428 & 1975 \\ 
  16 & Compositions & 252 & 1940 \\ 
  5 & Chemical apparatus and process disinfecting, deodorizing, preserving, or steril & 422 & 1978 \\ 
   \hline
  \hline
 \multicolumn{4}{|c|}{Classes of origins for Class 506}\\
   \hline
579 & Chemistry: molecular biology and microbiology & 435 & 1979 \\ 
  127 & Chemistry: analytical and immunological testing & 436 & 1982 \\ 
  69 & Chemical apparatus and process disinfecting, deodorizing, preserving, or steril & 422 & 1978 \\ 
   \hline
  \hline
 \multicolumn{4}{|c|}{Classes of origins for Class 706}\\
   \hline
966 & [NA] Information Processing System Organization & 395 & 1991 \\ 
  195 & [NA] Electrical Computers and Data Processing Systems & 364 & 1977 \\ 
  41 & Electrical transmission or interconnection systems & 307 & 1952 \\ 
   \hline
\end{tabular}
\caption{Number of patents pre-dating the creation of a class and reclassified into it, by class of origin; Only the three largest origin classes are shown, with their class number and date established.}
\label{table:reclass_origin}
\end{table}

Motivated by the study of \citet{erdi2013prediction} showing that the emergence of a new class (442) could have been predicted by citation clustering, we study class 442, ``Fabric (woven, knitted, or nonwoven textile or cloth, etc.)''. The class definition indicates that it is ``for woven, knitted, nonwoven, or felt article claimed as a fabric, having structural integrity resulting from forced interassociation of fibers, filaments, or strands, the forced interassociation resulting from processes such as weaving, knitting, needling hydroentangling, chemical coating or impregnation, autogenous bonding (\dots) or felting, but not articles such as paper, fiber-reinforced plastic matrix materials (PPR), or other fiber-reinforced materials (\dots)''. This class is ``an integral part of Class 428 [and as such it] incorporates all the definitions and rules as to subject matter of Class 428.'' The oldest patent reclassified in it was a patent by Charles Goodyear describing how applying caoutchouc to a woven cloth lead to a material with ``peculiar elasticity'' (US4099, 1845, no classification on the paper file). A first remark is that this class was relatively large at birth. Second, an overwhelming majority of patents came from the ``parent'' class 428. Our interpretation is that this is an example of an old branch of knowledge, textile, that due to continued development needs to be more finely defined to allow better classification and retrieval \-- note that the definition of 442 is not only about what the technologies are, but what they are not (paper and PPR).

Our second example is motivated by \possessivecite{kang2012science} qualitative study of the process of creation of an IPC class, to which the USPTO participated. \citet{kang2012science} describes that the process of class creation was initiated because of a high number of incoming patents on the subject matter. Her main conclusion is that disputes regarding class delineation were resolved by evaluating the size of the newly created category under certain definitions. Class 506, ``Combinatorial chemistry technology: method, library, apparatus'' includes in particular ``Methods specially adapted for identifying the exact nature (e.g., chemical structure, etc.) of a particular library member'' and ``Methods of screening libraries or subsets thereof for a desired activity or property (e.g., binding ability, etc.)''. The oldest reclassified patent is US3814732 (1974), ``modified solid supports for solid phase synthesis''. It claims polymeric hydrocarbon resins that are modified by the introduction of other compounds. It was reclassified from class 260, ``Chemistry of carbon compounds''. In contrast to 442 or 706 reviewed below, the reclassified patents are drawn relatively uniformly from several categories. Our interpretation is that this is an example of a mid-age technology (chemistry), which due to its interactions with other technologies (computers) develops a novel branch that is largely cross-cutting, but specific enough to warrant the creation of a new class.

Our last example is 706, ``Data processing \-- Artificial Intelligence'', which is a ``generic class for artificial intelligence type computers and digital data processing systems and corresponding data processing methods and products for emulation of intelligence (\dots); and including systems for reasoning with uncertainty (\dots), adaptive systems, machine learning systems, and artificial neural networks.''. We chose it because we possess at least some domain knowledge. The oldest reclassified AI patent is US3103648 (1963), which is an ``adaptive neuron having improved output'', nicely echoing the recent surge of interest in neural networks for machine learning (deep learning). It was originally classified in class 340, ``Communications:  electrical''. In contrast to the other two examples, we find that the two largest sources were classes that have since been abolished (we recovered the names of 395 and 364 from the ``1996 Index to the US patent classification''; their date established was available from the ``Date Established'' file documented in Section  \ref{section:data-construction}). Other classes with the ``Data processing'' header were created during the period, showing that the USPTO had to completely re-organize its computer-related classes around the turn of the millennium. Our interpretation is that this is an example of a highly novel technology, emerging within the broader context of the third and perhaps fourth industrial revolution. Because computers are relatively recent and general purpose, it is very difficult to create taxonomies with stable boundaries.

These three examples show strikingly different patterns of technological development and its associated classification volatility. An old branch of knowledge which is deepening (textile), a mid-age branch of knowledge that develops novel interactions with others (chemistry), and a new branch of knowledge (computers) for which classification officers strive to find useful organizational schemes. We acknowledge that these are only examples \-- presumably, some other examples of new classes would follow similar patterns, but other patterns may exists. We have found that about two thirds of post-1976 new classes have more than 90\% of their pre-birth (and post-1976) reclassified patents coming from a single origin (pre-existing class), suggesting that a form of ``branching'' or ``class splitting'' is fairly common, at least when looking at OR classes only. We do not want to put too much weight on these early results, which will have to be systematised, developed further using subclasses and multiple classifications, and, crucially, compared against results obtained using the IPC/CPC. We do think that such a systematic study of classification re-organizations would tell a fairly detailed story of the evolution of technology, but rather than embarking on such a detailed study here we propose to summarize most of what we have learned so far into a simple theoretical model.

\section{A simple model}
\label{section:model}

\begin{figure*}[p!]
\centering
\includegraphics[width=\textwidth]{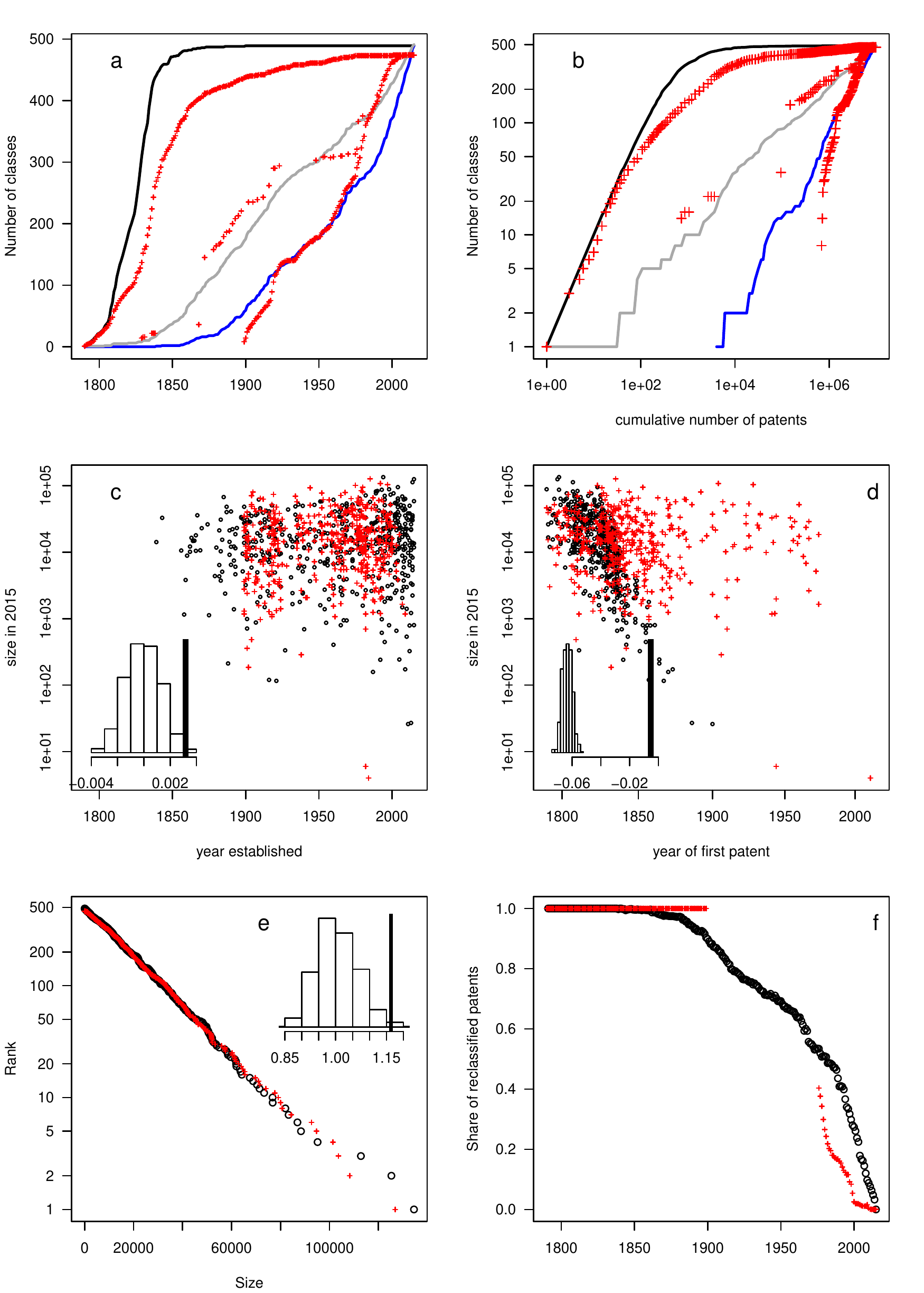}
\caption{Simulation results against empirical data (red crosses). See Section \ref{section:model} for details.}
\label{fig:simulationresults}
\end{figure*}

In this section, we propose a very simple model that reproduces several facts described above. As compared to other recent models for size distributions and Heaps' law in innovation systems \citep{tria2014dynamics,marengo2016arrival,lafondSDPC}, the key assumption that we will introduce is that classes are sometimes split and their items reclassified. We provide basic intuition instead of a rigorous discussion\footnote{For instance, we do not claim that the model \emph{in general} produces a certain type of pattern such as a lack of age-size relationship. We simply show that under a specific parametrisation taken from the empirical data (say $\sim$10 million patents, 500 classes, and a Heaps exponent of $0.38$), it produces patterns similar to the empirical data.}. 

Let us start with the well-known model of \citet{simon1955class}. A new patent arrives every period. The patent creates a new category with probability $\alpha$, otherwise it goes to an existing category which is chosen with probability proportional to its size. The former assumption is meaningful, because in reality the number of categories grows over time. The second assumption is meaningful too, because this ``preferential attachment''/``cumulative advantage'' is related to Gibrat's law: categories grow at a rate independent of their size, so that their probability of getting the next patent is proportional to their size.

There are three major problems with this model. First it gives the Yule-Simon distribution for the size distribution of classes. This is basically a power law so it has much fatter tails than the exponential law that we observe. In other words, it over predicts the number of very large categories by a large margin. Second, since older categories have more time to accumulate patents, it predicts a strong correlation between age and size. Third, since at each time step categories are created with probability $\alpha$ and patents are created with probability $1$, the relationship between the number of categories $\alpha t$ and the number of patents $t$ is linear instead of Heaps' constant elasticity relation.

A solution to make the size distribution exponential instead of power law is to change preferential attachment for uniform random attachment, that is to choose each category with equal probability. Besides the fact that this new assumption may seem less intuitive than Gibrat's law, this would not solve the second problem because it would still be the case that older categories accumulate more patents. The solution is to acknowledge that categories are not entities that are defined once and for all; instead, they are frequently split and their patents are reclassified.

We therefore turn to the model proposed by \citet{ijiri1975some}. It assumes that new categories are introduced over time by splitting existing ones. In its original form the model postulates a linear arrangement of stars and bars. Each star represents a patent, and bars materialize the classes. For instance, if there are 3 patents in class 1 and 1 patent in class 2, we have |***|*|. Now imagine that between any two symbols there is a space. At each period, we choose a space uniformly at random and fill it with either a bar (with probability $\alpha$) or a star (with complementary probability). When a star is added, it means that an existing category acquires a new patent. When a bar is added, it means that an existing category is split into two categories. It turns out that the resulting size-distribution is exponential, as desired. But before we can evaluate the age-size relationship, we need to decide how to measure the age of a category. To do this we propose to reformulate the model as follows.

We start with one patent in one category. At each period, we first select an existing category $j$ with probability proportional to its size $k_j$ and add one patent in it. Next, with probability $\alpha$ we create two novel categories by splitting the selected category uniformly at random; that is, we draw a number $s$ from a uniform distribution ranging from 1 to $k_j$. Next, each patent in $j$ is assigned to the new category 1 with the probability being $s/k_j$, or to the new category 2 otherwise. This procedure leads to a straightforward interpretation: the patents are \emph{reclassified} from $j$ to the first or the second new category. These two categories are \emph{established} at this step of the process, and since patents are created sequentially one by one, we also know the \emph{date of the first patent} of each new category. To give a date in calendar years to patents and categories, we can simply use the dates of the real patents.

Since $\alpha$ is constant, as in Simon's original model, we are left with the third problem (Heaps' power law is violated). We propose to make $\alpha$ time dependent to solve this issue\footnote{An interesting alternative (instead of using the parameter $\alpha$) would be to model separately the process by which the number of patents grow and patent classification officers split categories.}. Denoting the number of categories by $C_t$ and the number of patents by $t$ (since there is exactly one new patent per period), we want to have $C_t = C_0 t^b$ (Heap's law). This means that $C_t$ should grow at a per period rate of $dC_t/dt=C_0 b t^{b-1}$. Since we have measured $b \approx 0.378$ and we want the number of categories to be 474 when the number of patents is 9,847,315, we can calculate $C_0=C_t/t^b=1.07$. This gives $\alpha_t = 1.07 \times 0.378 \hspace{1mm} t^{0.378-1}$, which we take to be 1 when $t=1$.\footnote{There is a small inconsistency arising because the model is about primary classification only, but the historical number of classes and Heaps' law are measured using all classes, because we could not differentiate cross-reference classes in historical data. Another point of detail is that we could have used the estimated $C_0=0.17$ instead of the calculated one. These details do not fundamentally change our point.}

Note how parsimonious the model is: its only inputs are the current number of patents and categories, and the Heaps' exponent. Here we do not attempt to study it rigorously. We provide simulation results under specific parameter values. Fig. \ref{fig:simulationresults} shows the outcome of a single simulation run (black dots and lines), compared to empirical data (red crosses). 

The first pair of panels (a and b) shows the same (empirical) data as Fig. \ref{fig:Nclasses} and \ref{fig:Heaps} using red crosses. The results from the simulations are the curves. The simulation reproduces Heaps' law well, by direct construction (the grey middle curve on panel b). But it also reproduces fairly well the evolution of the reconstructed number of classes, both the one based on the ``date of first patent'' and the one based on the ``dates established'', and both against calendar time (years) and against the cumulative number of patents.

The second pair of panels (c and d) show the age-size relationships, with the same empirical data as in Fig. \ref{fig:agesize}. Panel c shows that the model seems to produce categories whose sizes are \emph{not} strongly correlated with the year in which they were established, as in the empirical data. However, in panel d, in our model there is a fairly strong negative correlation between size and the year of the first patent and this correlation is absent (or is much weaker) in the empirical data. These results for one single run are confirmed by Monte Carlo simulations. We ran the model 500 times and recorded the estimated coefficient of a simple linear regression between the log of size and each measure of age. The insets show the distribution of the estimated coefficients, with a vertical line showing the coefficient estimated on the empirical data. 

The next panel (e) shows the size distribution in a rank-size form, as in Fig. \ref{fig:ranksize}. As expected, the model reproduces this feature of the empirical data fairly well. However the empirical data is not exactly exponential and may be slightly better fitted by a negative binomial model (which has one more parameter and recovers the exponential when its shape parameter equals one). The top right histogram shows the distribution of the estimated negative binomial shape parameter. The empirical value departs only slightly from the Monte Carlo distribution.

Finally, the last panel (f) shows the evolution of the share of reclassified patents, with the empirical data from Fig. \ref{fig:sharereclass} augmented by values of 1 between 1790 and 1899 (since no current categories existed prior to 1899, all patents have been reclassified). Here again, the model reproduces fairly well the empirical pattern. All or almost all patents from early years have been reclassified, and the share is falling over time. That said, for recent years (post 1976), the specific shape of the curve is different.

Overall, we think that given its simplicity the model reproduces a surprisingly high number of empirical facts. It allows us to understand the differences between the different patterns of growth of the reconstructed and historical number of classes. Without a built-in reclassification process it would not have been possible to match all these empirical facts \--- if only because without reclassification historical and reconstructed evolution coincide. This shows how important it is to consider reclassification when we look at the mesoscale evolution of the patent system. On the other hand, much more could be done to make the model more interesting and realistic, for instance by also modelling subclasses and requiring that reclassification takes place within a certain distance.

\section{Conclusion}

In this paper, we have presented a quantitative history of the evolution of the main patent classes within the U.S. Patent Classification System. Our main finding is that the USPCS incurred regular and important changes. For academic researchers, these changes may be perceived as a source of problems, because this suggests that it may not always be legitimate to think that a given patent belongs to one and the same category forever. This means that results obtained using the current classification system may change in the future, when using a different classification system, and even if the very same set of patent is considered.

That said, we do not think the effect would be strong. Besides, using the current classification system is still often the best thing to do because of its consistency. Our point here is not to critique the use of the current classification, but to argue that historical changes to the classification system itself contain interesting information that has not been exploited.

Our first result is that different methods to compute the growth of the number of classes give widely different results, establishing that the changes to the classification system are very important. Our second result suggests that we do not see very large categories in empirical data because categories are regularly split, leading to an exponential size distribution with no relationship between the age and size of a category. Our third result is that reclassification data contains useful information to understand technological evolution. Our fourth result is that a very simple model that can explain many of the observed patterns needs to include the splitting of classes and the reclassification of patents. Taken together, these results show that it is both necessary and interesting to understand the evolution of classification systems. 

An important limitation of our study is that it is highly limited in scope: we study the US, at the class level, using main classifications only. A contrasting example we have found is the French patent classification of 1853, which contained 20 groups, was revised multiple times in the $19^{th}$ century but while subclasses were added it kept a total of 20 classes even in the ``modern'' classification of 1904. Similarly, while direct comparison is difficult, our preliminary exploration of other classification systems, such as the IPC and CPC, suggests that they do not feature the same size distribution, perhaps pointing to a different mode of evolution than the one proposed in our model.

We believe that our findings are interesting for all researchers working with economic and technological classifications, because we characterized quantitatively the volatility of the patent classification system. We do not know whether they are unstable because collective representations of technological artefacts are context-dependent, or because as more items are introduced and resources invested in classifying them appropriately, collective discovery of the ``true'' mesoscale partition takes place. But clearly, when interpreting the results which rely upon a static snapshot of a classification system, one should bear in mind that classification systems are dynamic.

A case in point is the use of technological classes to produce forecasts: how can we predict the evolution of a given class or set of classes several decades ahead, when we know these classes might not even exist in the future? In this paper, we are not proposing a solution to this forecasting issue \-- only raising conceptual problems that classification system changes pose. Further, even if we consider that today's categorization will not change, a subtle issue arises in the production of correct forecasting models. To see this consider developing a time series model describing the growth of some particular classes. To test the forecasting ability of the model, one should perform out-of-sample tests, as e.g. \citet{farmer2016predictable} did for technology performance time series. Part of the past data is used to predict more recent data, and the data which is not used for estimation is compared to the forecasts. Now, note that when we use the current classification, we effectively use data from the present; that is, the delineation of categories for past patents uses knowledge from the present, and it is therefore not entirely valid to evaluate forecasts (there is ``data snooping'' in the sense that one uses knowledge of the future to predict the future).

Classification system changes pose serious problems for forecasting but may also bring opportunities: if classification changes reflect technological change then one can in principle construct quantitative theories of that change. Since the patterns described here could be roughly understood using an extremely simple model, it may be possible to make useful forecasts with more detailed models and data, for instance predicting new classes \citep{erdi2013prediction,kyebambe2017forecasting}. This could be useful because patent classification changes are more frequent than changes to other classification systems such as industries, products and occupations. An interesting avenue for future research would be to use the changes of the patent classification system to predict the changes of industry and occupation classification systems, thus predicting the types of jobs of the future.

Beyond innovation studies, with the rising availability of very large datasets, digitized and carefully recorded classifications and classification changes will become available. It will be possible to explore classifications as an evolving network and track the splitting, merging, birth and death of categories. This is an exciting new area of research, but the big data that we will accumulate will only (or mostly) cover recent years. This makes historical studies such as the present one all the more important.

\bibliographystyle{agsm}
\bibliography{bib}

\end{document}